# Second Order Scattering Models of Elastic Waves in Heterogeneous Polycrystalline Materials

Gaofeng Sha[1,2]

1. Department of Mechanical and Aerospace Engineering, The Ohio State University

2. Department of Materials Science and Engineering, The Ohio State University

Email: sha.34@buckeyemail.osu.edu

**Abstract**: A general second order attenuation (SOA) model is proposed to predict the elastic wave attenuation and phase velocity dispersion in heterogeneous polycrystalline media. It is valid for statistically isotropic aggregates with triclinic crystals of equiaxed shape and is equivalent to well-known Stanke&Kino model when applied to cubic polycrystals. Moreover, an approximation form of the general SOA model is obtained to improve computational efficiency but retain adequate accuracy. Further comparison between the SOA model and the approximated second order attenuation (ASOA) model indicates the ASOA model has reasonable agreement with the SOA model both on attenuation and phase velocity. Additional computational examples show the ASOA model has better performance than Karal&Keller model. Thus, this theoretical study provides effective approaches for modeling of acoustic attenuation in heterogeneous polycrystalline materials for whole frequency range, including the Rayleigh region, the stochastic region and the geometric region. It will shed light on the practical development of ultrasonic characterization of polycrystalline metals.

# 1. Introduction

Modeling of wave scattering in polycrystalline materials has been ongoing for a century. This study can date back to 1940s [1] and later was renewed by Merkulov[2], though only asymptotic attenuation models were available that time. A review of early attenuation modeling and validation can be found in Ref.[3]. Another model derived based on scalar wave but applicable for electromagnetic and acoustic wave is Karal&Keller Model [4], and it has been extended to elastic wave in hexagonal polycrystals[5]. One remarkable scattering model is the unified theory developed by Stanke & Kino[6], which is applicable to polycrystal aggregate with cubic crystallites of equiaxed shape. Alternatively, Weaver [7] derived acoustic attenuation in polycrystals using Dyson equation and scattering coefficient in terms of Bethe-Salpeter equation and provided explicit attenuation coefficients for cubic polycrystalline materials of equiaxed grains under the frame of Born approximation. It is worth mentioning that Weaver type model has been extended to macroscopically isotropic aggregates of triclinic[8], [9] grains while Li&Rokhlin [10] further extended Weaver type model to textured polycrystal aggregates of triclinic grains in ellipsoidal shape. However, those general models[8]–[10] except for second order models[6], [11] break at high frequency due to Born approximation and may have large discrepancy with second order model for grains with large anisotropy.

Other studies have thrived to extend Stanke&Kino (second order) model to more general cases, like polycrystals with ellipsoidal grains [12] and polycrystals with realistic two-point correlation function[11], [13]. Another kind of second order models [14], [15] adopted spectral function approach and are valid for whole frequency range since those models account for some multiple scattering events. These models [14], [15] account for statistically isotropic aggregates of monoclinic polycrystalline materials with ellipsoidal grains. However, no second order model has been reported for statistically isotropic aggregates of triclinic crystallites and existing second order model[14] for low symmetry grains are computationally expensive.

This study proposes a second order attenuation (SOA) model for statistically isotropic aggregates with triclinic grains of equiaxed shape, and it can produce both attenuation and phase velocity for the whole frequency range. This SOA model is further simplified into an approximated form, ASOA model, to achieve higher computational efficiency and better accuracy (better than far field scattering model in [16]).

In the following sections, theoretical background about attenuation modeling will be addressed first in Section 2. It is followed by the SOA model and the ASOA model in Section 3. In section 4, computation results will be discussed, and different models will be compared. Finally, conclusions are provided.

## 2. Theoretical background

This section will review the theoretical background about modeling of elastic wave attenuation in polycrystalline materials.

### 2.1 Dyson equation and mass operator

As stated in Refs.[6], [7], there are three methods for attenuation modeling, perturbation theory[6], Dyson equation[7] and spectral function method [14]. Here we follow Weaver's approach[7], the mean Green function response $\langle G_{k\alpha}(\mathbf{X},\mathbf{X}')\rangle$ in the heterogeneous polycrystals can be obtained by Dyson equation:

$$\langle G_{k\alpha}(\mathbf{X},\mathbf{X}')\rangle = G_{k\alpha}^0(\mathbf{X},\mathbf{X}') + \iint G_{k\beta}^0(\mathbf{X},\mathbf{Y})M_{\beta j}(\mathbf{Y},\mathbf{Z})\langle G_{j\alpha}(\mathbf{Z},\mathbf{X}')\rangle d^3y d^3z \qquad (1)$$

where $G_{k\alpha}^0(\mathbf{X},\mathbf{X}')$ is the Green function of homogenized medium or reference medium and $M_{\beta j}(\mathbf{Y},\mathbf{Z})$ is the mass operator or self-energy accounting for all possible scattering events [17]. The mass operator could be expressed in diagrammatic form including infinite series[18]. However, the exact equation for mass operator is difficult to obtain but finite order approximation is given in references[7], [18]. Under the weak scattering assumption $|\delta c_{ijkl}|/C_{ijkl}^0 \ll 1$ ($C_{ijkl}^0$ is effective elastic constants of the polycrystals aggregate and $\delta c_{ijkl}$ is the local fluctuation of elastic constants due to random crystallite orientation), mass operator after first order smoothing approximation (FOSA) is [19]:

$$M_{\beta j}(\mathbf{Y},\mathbf{Z}) \approx \left\langle \frac{\partial}{\partial Y_\alpha} \delta c_{\alpha\beta\gamma\delta}(\mathbf{Y}) \frac{\partial}{\partial Y_\delta} G_{\gamma q}^0(\mathbf{Y},\mathbf{Z}) \frac{\partial}{\partial Z_i} \delta c_{ijql}(\mathbf{Z}) \frac{\partial}{\partial Z_l} \right\rangle \qquad (2)$$

where $G_{\gamma q}^0$ is the Green function for homogenized medium and $\delta c_{ijkl}$ is the spatial variation of elastic constants. After spatial Fourier transform of the double convolution, Dyson equation could be simplified as (more details could be found in Refs.[14] and [16]):

$$\langle \mathbf{G}(\mathbf{k}) \rangle = \left[ \mathbf{G}^0(\mathbf{k})^{-1} - \mathbf{M}(\mathbf{k}) \right]^{-1} \tag{3}$$

where $\mathbf{k}$ is the perturbed wavenumber of an elastic wave propagating in heterogeneous polycrystalline media and the Green tensor in the reference medium is[10], [16]:

$$\mathbf{G}^0(\mathbf{k}) = \sum_{M=1}^{3} \frac{\mathbf{u}^M \otimes \mathbf{u}^M}{\omega^2 - k^2 V_M^2} = \sum_{M=1}^{3} g_M^o \mathbf{u}^M \otimes \mathbf{u}^M \tag{4}$$

where $\otimes$ stands for dyadic produce, $\mathbf{u}^M$ denotes a polarization vector of wave mode $M$ (either a longitudinal wave L or transverse wave T), $V_M$ means velocity of mode $M$, and $\omega = k_{0M} V_M$ is the angular frequency.

The explicit equation for mass operator relevant to the covariance of elastic constants and two-point correlation has been reported in several references[10], [16], [19], [20]. Here it is given:

$$\mathbf{M}_{\beta j}(\mathbf{k}) = k_\alpha k_l \int G_{\gamma q}^0(\mathbf{k}^S) k_\delta^S k_i^S \langle \delta c_{ijql} \delta c_{\alpha\beta\gamma\delta} \rangle W(\mathbf{k} - \mathbf{k}^S) d^3 k^S \tag{5}$$

The mass operator could be decomposed into a summation of three orthogonal tensors and so is the mean Green tensor[10], [19]. Therefore, one could obtain:

$$\mathbf{M}(\mathbf{k},\omega) = \sum_{M=1}^{3} m_M \mathbf{u}^M \otimes \mathbf{u}^M \quad ; \quad \langle \mathbf{G}(\mathbf{k},\omega) \rangle = \sum_{M=1}^{3} g_M \mathbf{u}^M \otimes \mathbf{u}^M \tag{6}$$

The three components of the mass operator above are $m_M(\mathbf{k})$ [16]:

$$m_M(\mathbf{k}) = \sum_{N=1}^{3} m_{M \to N} = \frac{k^2(\mathbf{p})}{\rho^2} \sum_{N=1}^{3} \int g_0^N(\mathbf{k}^{S,N}) IP_{M \to N}(\mathbf{p},\mathbf{s}) k_S^2 W(\mathbf{k} - \mathbf{k}^{S,N}) d^3 k^{S,N}, \tag{7}$$

where $\rho$ is material density, $M$ and $N$ stand for the modes of an incident wave and a scattered wave respectively (longitudinal wave or two transverse waves), and inner product $IP_{M \to N}(\mathbf{p},\mathbf{s}) = \langle \delta c_{ijql} \delta c_{\alpha\beta\gamma\delta} \rangle u_\beta u_q p_\alpha p_l s_i s_\delta v_\gamma v_j$ associated with the wave propagation direction $\mathbf{p}$ (polarization vector $\mathbf{u}$) and the scattering direction $\mathbf{s}$ (polarization vector $\mathbf{v}$). The mean response $g_M(\mathbf{k})$ (see Eq.6) of an incident wave (mode M) in the polycrystals aggregate is[7], [10], [19]:

$$g_M(\mathbf{k}) = \left[ g_M^o(\mathbf{k})^{-1} - m_M(\mathbf{k}) \right]^{-1} = \left[ \omega^2 - k^2 V_M^2 - m_M(\mathbf{k}) \right]^{-1} \tag{8}$$

Since the denominate of mean Green function should equal to zero to provide the excitations of the polycrystalline system, the dispersion equation is expressed as[10], [19]:

$$\omega^2 - k^2 V_M^2 - m_M(\mathbf{k}) = 0 \tag{9}$$

where $k = \text{Re}\,k + i\alpha$ is the complex wave number in the perturbed medium. The real part could provide frequency dependent velocity while the imaginary corresponds to wave attenuation. Note that Calvet&Margerin [14] utilized the spectral function instead of Eq.(9) to determine the amplitude of different wave modes.

## 2.2 Elastic constants covariance and two-point correlation function

The mass operator aforementioned is related to two-point correlation (TPC) function [6], [7], $\delta c_{ijkl}(\mathbf{X})\delta c_{\alpha\beta\gamma\delta}(\mathbf{X}')$, a covariance of elastic constants fluctuation from two points $\mathbf{X}$ and $\mathbf{X}'$. Since we are solely interested in the mean wave response in heterogeneous polycrystals, the ensemble average of this covariance is critical. From statistics of numerous grains, the ensemble average of elastic covariance could be further decomposed into two parts: the volumetric average of elastic constants fluctuation covariance and geometric two-point correlation (TPC) function [6], [7], [21]. The statistical two-point correlation function is:

$$\langle \delta c_{ijkl}(\mathbf{X})\delta c_{\alpha\beta\gamma\delta}(\mathbf{X}')\rangle = \langle \delta c_{ijkl}\delta c_{\alpha\beta\gamma\delta}\rangle w(\mathbf{X}-\mathbf{X}') \tag{10}$$

where $\langle \delta c_{ijkl}\delta c_{\alpha\beta\gamma\delta}\rangle$ is the ensemble average of elastic constants covariance and $w(\mathbf{X}-\mathbf{X}')$ geometric two-point correlation function. It could be explicitly calculated through single crystal elastic constants and orientation distribution function (ODF) through the following equation [10], [11], [22], [23]:

$$\langle \delta c_{ijkl}\delta c_{\alpha\beta\gamma\delta}\rangle = \langle c_{ijkl}c_{\alpha\beta\gamma\delta}\rangle - \langle c_{ijkl}\rangle\langle c_{\alpha\beta\gamma\delta}\rangle, \tag{11a}$$

$$\langle c_{ijkl}c_{\alpha\beta\gamma\delta}\rangle = \frac{1}{8\pi^2}\int_0^\pi \int_0^{2\pi}\int_0^{2\pi} c_{ijkl}(\theta,\phi,\zeta)c_{\alpha\beta\gamma\delta}(\theta,\phi,\zeta)\sin\theta d\phi d\zeta d\theta \tag{11b}$$

$$C_{ijkl}^0 = \langle c_{ijkl}\rangle = \frac{1}{8\pi^2}\int_0^\pi \int_0^{2\pi}\int_0^{2\pi} c_{ijkl}(\theta,\phi,\zeta)\sin\theta d\phi d\zeta d\theta \tag{11c}$$

where $C^0_{ijkl}$ are the effective elastic moduli of the polycrystals aggregate by Voigt average[8], $c_{ijkl}(\theta,\phi,\zeta)$ is the elastic constants for a single crystal after the rotation by three Euler angles $\theta,\phi,\zeta$ [8], [10] and ODF is $\frac{1}{8\pi^2}$ for randomly orientated grains [8], [10].

Since the polycrystals aggregate is macroscopically isotropic, the geometric two-point correlation function $w(\mathbf{X}-\mathbf{X}')$ is solely dependent on the distance between two points so that this relation holds $w(\mathbf{X}-\mathbf{X}') = w(r)$, where $r = |\mathbf{X}-\mathbf{X}'|$. Furthermore, following Refs.[6]–[8], [14], this paper considers polycrystals microstructures with uniform equiaxed grains, which corresponds to an exponential form two-point correlation function [7], [24]:

$$w(r) = \exp(-r/a) \tag{12}$$

where a is the mean intercept length on material cross section [24]. The spatial Fourier transform of such a two-point correlation function is:

$$W\left(\mathbf{k}-\mathbf{k}^{S,N}\right) = \frac{a_X a_Y a_Z}{\pi^2 \left[1 + a^2\left(k^2 + (k^{S,N})^2 - 2kk^{S,N}\cos\theta\right)\right]^2} \tag{13}$$

where $\mathbf{k}$ means incident wavenumber, $\mathbf{k}^{S,N}$ scattered wavenumber and $\theta$ is the angle between wave propagation direction $\mathbf{p}$ and scattering direction $\mathbf{s}$. One should note that other forms of two-point correlation function corresponding to different grain size distributions could be found in Refs.[11], [25], [26]. However, for simplicity this paper only accounts for exponential form TPC functions.

## 3. Second order attenuation model and approximated model

This section will firstly address the second order attenuation model derived based on Refs. [14] and [16]. Then an approximated second order attenuation model is obtained to improve computation efficiency.

### 3.1 Second order attenuation model

The dispersion equation is given, however, the complete equation for mass operator is not determined yet. In Eq.(6), the expression of Green function for one wave mode is [14]:

$$g_0^N(k^{S,N}) = \frac{1}{(\omega+i\varepsilon)^2 - (k^{S,N}V_N)^2} = P.V.\left(\frac{1}{\omega^2 - (k^{S,N}V_N)^2}\right) - i\pi\delta[\omega^2 - (k^{S,N}V_N)^2] \quad (14)$$

where superscript N donates the scattered wave mode (), $V_N$ is velocity, $\varepsilon$ is a positive infinitesimal number, $P.V.$ stands for Cauchy principal value and $\delta$ is the Delta function. After substitute the Green Function Eq. (14) into the dispersion equation (9), one could obtain the second order attenuation model. The dispersion equation for a longitudinal incidence wave could be rewritten as:

$$k_{0L}^2 - k^2 - m_L(k)/V_{0L}^2 = 0 . \quad (15)$$

And the mass operator is split into two scattering components due to mode conversion [7]:

$$m_L(k) = m_{L \to L}(k) + m_{L \to T}(k) \quad (16)$$

$$m_{L \to L}(k) = \frac{2k^2 a^3 k_{0L}^3}{\pi \rho^2 V_{0L}^2} P.V. \int_0^\infty \frac{s^4}{1-s^2} dS \int_{-1}^{+1} \frac{A_{LL} + B_{LL}x^2 + C_{LL}x^4}{\left[(1+k^2a^2 + s^2 k_{0L}^2 a^2) - 2skk_{0L}a^2 x\right]^2} dx$$

$$-i\frac{k^2 a^3 k_{0L}^3}{\rho^2 V_{0L}^2} \int_{-1}^{+1} \frac{A_{LL} + B_{LL}x^2 + C_{LL}x^4}{\left[(1+k^2a^2 + k_{0L}^2 a^2) - 2kk_{0L}a^2 x\right]^2} dx$$

$$m_{L \to T}(k) = \frac{2k^2 a^3 k_{0T}^3}{\pi \rho^2 V_{0T}^2} P.V. \int_0^\infty \frac{s^4}{1-s^2} dS \int_{-1}^{+1} \frac{A_{LT} + B_{LT}x^2 + C_{LT}x^4}{\left[(1+k^2a^2 + s^2 k_{0T}^2 a^2) - 2skk_{0L}a^2 x\right]^2} dx$$

$$-i\frac{k^2 k_{0T}^3 a^3}{\rho^2 V_{0T}^2} \int_{-1}^{+1} \frac{A_{LT} + B_{LT}x^2 + C_{LT}x^4}{\left[(1+k^2a^2 + k_{0T}^2 a^2) - 2kk_{0T}a^2 x\right]^2} dx$$

where $L \to L, L \to T$ donates L-L mode scattering and L-T mode scattering, $x = \cos\theta$, and Voigt velocities calculated from effective moduli of polycrystals aggregates are:

$$\langle c_{11} \rangle = \rho V_{0L}^2 = \frac{1}{5}(c_{11} + c_{22} + c_{33}) + \frac{2}{15}(c_{12} + c_{23} + c_{31}) + \frac{4}{15}(c_{44} + c_{55} + c_{66})$$

$$\langle c_{44} \rangle = \rho V_{0T}^2 = \frac{1}{15}(c_{11} + c_{22} + c_{33}) - \frac{1}{15}(c_{12} + c_{23} + c_{31}) + \frac{1}{5}(c_{44} + c_{55} + c_{66}) \quad (17)$$

The inner product coefficients $A_{LL}, B_{LL}, C_{LL}$ $A_{LT}, B_{LT}, C_{LT}$ for aggregates of triclinic grains are given in Ref.[16] and for higher symmetry crystallites such as monoclinic and hexagonal symmetry classes could be found in Ref.[14].

Similarly, the dispersion equation for a transverse incident wave is:

$$k_{0T}^2 - k^2 - m_T(k)/V_{0T}^2 = 0 \qquad (18)$$

where the mass operator is written as:

$$m_T(k) = m_{T \to L}(k) + m_{T \to T}(k) \qquad (19a)$$

$$m_{T \to T}(k) = \frac{k^2 a^3 k_{0T}^3}{\pi \rho^2 V_{0T}^2} P.V. \int_0^\infty \frac{s^4}{1-s^2} ds \int_{-1}^{+1} \frac{A_{TT} + B_{TT}x^2 + C_{TT}x^4}{\left[(1+k^2a^2+s^2k_{0T}^2a^2)-2ksk_{0T}a^2x\right]^2} dx$$

$$-i\frac{k^2 k_{0T}^3 a^3}{2\rho^2 V_{0T}^2} \int_{-1}^{+1} \frac{A_{TT} + B_{TT}x^2 + C_{TT}x^4}{\left[(1+k^2a^2+k_{0T}^2a^2)-2kk_{0T}a^2x\right]^2} dx \qquad (19b)$$

$$m_{T \to L}(k) = \frac{k^2 a^3 k_{0L}^3}{\pi \rho^2 V_{0L}^2} P.V. \int_0^\infty \frac{s^4}{1-s^2} ds \int_{-1}^{+1} \frac{A_{LT} + B_{LT}x^2 + C_{LT}x^4}{\left[(1+k^2a^2+s^2k_{0L}^2a^2)-2ksk_{0L}a^2x\right]^2} dx$$

$$-i\frac{k^2 k_{0L}^3 a^3}{2\rho^2 V_{0L}^2} \int_{-1}^{+1} \frac{A_{LT} + B_{LT}x^2 + C_{LT}x^4}{\left[(1+k^2a^2+k_{0L}^2a^2)-2kk_{0L}a^2x\right]^2} dx \qquad (19c)$$

The two scattering modes T-T and T-L take place during transverse wave propagation in the polycrystals. Additionally, inner product coefficients $A, B, C$ for aggregates of triclinic grains have been reported in Refs.[16], [27]. All the coefficients are listed by covariances [16]:

$$A_{LL} = \langle \delta c_{13} \delta c_{13} \rangle, \quad B_{LL} = -2\langle \delta c_{13} \delta c_{13} \rangle + 4\langle \delta c_{15} \delta c_{15} \rangle + 2\langle \delta c_{13} \delta c_{33} \rangle;$$

$$C_{LL} = \langle \delta c_{33} \delta c_{33} \rangle + \langle \delta c_{13} \delta c_{13} \rangle - 4\langle \delta c_{15} \delta c_{15} \rangle - 2\langle \delta c_{13} \delta c_{33} \rangle;$$

$$A_{TT} = \langle \delta c_{44} \delta c_{44} \rangle + 3\langle \delta c_{45} \delta c_{45} \rangle, \quad B_{TT} = \langle \delta c_{44} \delta c_{44} \rangle - \langle \delta c_{45} \delta c_{45} \rangle - C_{LL}; \qquad (20)$$

$$A_{TL} = \langle \delta c_{14} \delta c_{14} \rangle + \langle \delta c_{15} \delta c_{15} \rangle, \quad B_{TL} = \langle \delta c_{15} \delta c_{15} \rangle - \langle \delta c_{14} \delta c_{14} \rangle + C_{LL};$$

$$C_{LL} = C_{TT} = -C_{TL}.$$

Therefore, the second order attenuation models both for longitudinal wave and transverse wave have been obtained. Their Born approximation forms could produce the same results as Kube&Turner model [8]. Also, they are equivalent to S&K model [6] when applied to cubic polycrystalline materials, which will be discussed below.

**3.2 Approximated Second Order Attenuation (ASOA) Model**

It is worthy to mention the general second order attenuation model in section 3.1 complicates the numerical calculation due to Cauchy principle value and computation time is much longer than classic Stanke&Kino model [6]. Moreover, the attenuation at Rayleigh region could not be determined accurately due to infinite upper limit of Cauchy principle value. Instead, we are looking for one approximated form of second order attenuation model in this section.

After many trials, it is found that one reasonable approach to simplify the dispersion equation is to apply asymptotic method at stochastic limit ($k_M a \gg 1$) and obtain the Cauchy principle value analytically. The inner product also could be factored out from integration like Ref.[16]. After some simplification, the approximated second order attenuation equation for longitudinal wave is:

$$k^2 = k_{0L}^2 + 4\left[Q'_{LL} + Q_{LT}\right]k^2 + \frac{4Q''_{LL}k^2 k_{0L}^2 a^2}{1 - 2iak_{0L} + a^2(k^2 - k_{0L}^2)} + \frac{4Q_{TL}k^2 k_{0T}^2 a^2}{1 - 2iak_{0T} + a^2(k^2 - k_{0T}^2)} \tag{21}$$

where the scattering factors related to materials elastic property are defined as:

$$Q'_{LL} = \frac{A_{LL} + B_{LL}/3 + C_{LL}/5}{4\rho^2 V_{0L}^4}; Q''_{LL} = \frac{A_{LL} + B_{LL} + C_{LL}}{4\rho^2 V_{0L}^4}$$

$$Q_{LT} = \frac{A_{LT} + B_{LT}/3 + C_{LT}/5}{4\rho^2 V_{0T}^2 V_{0L}^2} \tag{22}$$

Similarly, the approximated attenuation model for transverse wave is:

$$k^2 = k_{0T}^2 + 4\left[Q'_{TT} + Q_{TL}\right]k^2 + \frac{4Q''_{TT}k^2 k_{0T}^2 a^2}{1 - 2iak_{0T} + a^2(k^2 - k_{0T}^2)} + \frac{4Q_{TL}k^2 k_{0L}^2 a^2}{1 - 2iak_{0L} + a^2(k^2 - k_{0L}^2)} \tag{23}$$

where the scattering factors for transverse incident wave are:

$$Q'_{TT} = \frac{A_{TT} + B_{TT}/3 + C_{TT}/5}{8\rho^2 V_{0T}^4}; Q''_{TT} = \frac{A_{TT} + B_{TT} + C_{TT}}{8\rho^2 V_{0T}^4}$$
$$Q_{TL} = \frac{Q_{LT}}{2} = \frac{A_{LT} + B_{LT}/3 + C_{LT}/5}{8\rho^2 V_{0T}^2 V_{0L}^2}$$
(24)

One should note that $Q''_{LL}, Q_{LT}$ in Eq. (22) and $Q''_{TT}, Q_{TL}$ in Eq. (24) are identical to those defined in Ref.[16]; however, dispersion equations (21) and (23) are different from the far field attenuation model [16]. The far field attenuation model fails to produce correct phase velocity and additional correction is needed [16], but ASOA model can produce correct phase velocity (see Section 4.2). Dispersion equations (21) and (23) have the similar form as Karal&Keller model [4], [5]; however, Karal&Keller model is limited to scalar wave while ASOA model considers mode conversion during wave scattering.

## 4. Computation Results and Discussion

This section will describe computational examples using SOA and ASOA models and comparison between different models.

### 4.1 Comparison of second order attenuation model with Stanke&Kino model

As stated in Section 3, the second order attenuation model is a generalization of classic S&K model [6]. To prove this point, one cubic iron polycrystal with equiaxed grain in Ref. [6] is employed to compare SOA model with S&K model [6]. The single crystal elastic constants for iron are taken from Table #1 in Ref. [6]. Attenuation and phase velocity are calculated via second order attenuation model both for longitudinal wave and transverse wave, while these for S&K model are calculated through dispersion equations listed in Ref.[6]. All the numerical computations are implemented by Fortran code in this entire work. Also, Fortran library provides a standard subroutine that can calculates Cauchy principle value numerically.

After the calculations are completed, results for longitudinal wave and transverse wave are compared separately in Figure 1 and Figure 2. One should note that only dominant roots for attenuation models are present in this study, though two physical roots exist as stated by Calvet&Margerin [14]. The subdominant root corresponds to an evanescent wave and it has no practical application. In Figure 1a and Figure 2a dimensionless attenuation and frequency are used while relative velocity difference is plotted in Figure 1b and Figure 2b. Figure 1 and Figure 2 indicate two models are overlapped in whole frequency range, including Rayleigh regime, transition region, stochastic regime and geometric region. Quantitively comparisons shows that all results from two models are identical up to five digits (six digits when $2ka > 6.7$), which results in a relative difference comparable to the computation error set in Fortran code. Good agreement of higher digits between those two models is achievable; however, the Cauchy principle value must be evaluated more carefully and setting of a smaller computation error is necessary. It requires excessive effort and significant computational time; thus, it is out of the scope of this work.

The comparisons manifest that Stanke&Kino model and second order attenuation model presented in this work are equivalent for modeling of elastic wave attenuation in cubic polycrystalline materials. Nonetheless, the SOA model in this study is applicable to polycrystals aggregates of triclinic grains.

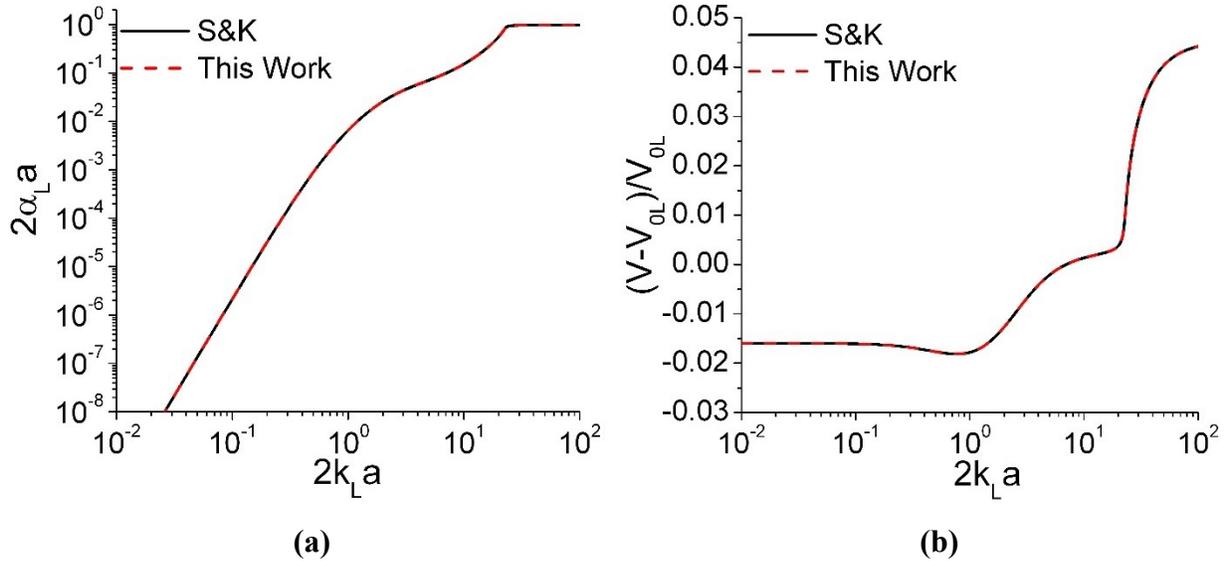

(a)                    (b)

Figure 1 Comparison of Stanke&Kino model with second order attenuation model (this work) for a cubic iron polycrystal with equiaxed grains: (a) longitudinal attenuation (b) longitudinal phase velocity V. Only dominant root is plotted, though two roots exist as demonstrated by [14]; frequency and attenuation are normalized by intercept length (grain radius) a.

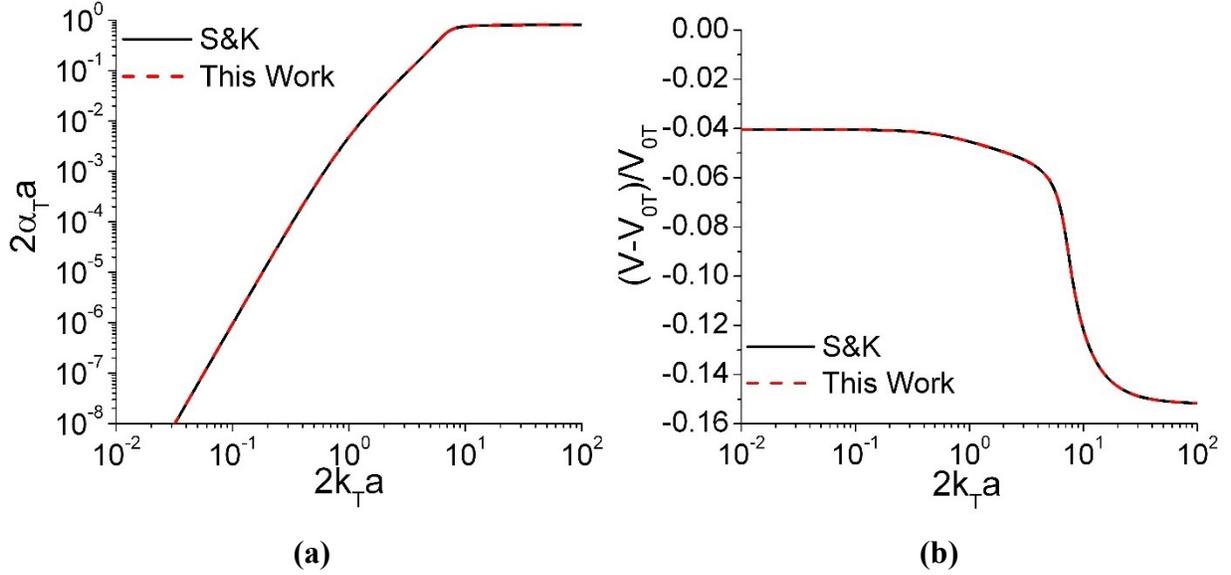

Figure 2 Comparison of second order attenuation model (this work) with Stanke&Kino model for a cubic iron polycrystal with equiaxed grains: (a) transverse attenuation (b) transverse phase velocity V. Only dominant root is provided, though two roots exist as demonstrated by [14]; frequency and attenuation are normalized by mean intercept length (grain radius) a.

**4.2 Comparison of SOA model with ASOA model**

Cubic iron and triclinic albite[16] are employed to verify the ASOA model developed in Section 3.2 by comparison with the SOA model. The elastic constants of single crystal iron are taken from Ref.[6] and those for albite are from Ref.[16]. Longitudinal attenuation and phase velocity from two models for these two materials are compared in Figure 3 and Figure 4. It is worthy to mention that approximated attenuation models are tens of times faster than the SOA model.

In Figure 3 attenuation curves from second order attenuation model and approximated model have reasonable agreement on both materials. The difference between two models are subtle; however, quantitative analysis indicates that maximum relative difference between these two models is 7% on cubic iron and 10.8% on albite. Another intriguing finding in Figure 3 is that albite has much short stochastic region than iron. It can be explained by the fact that wave scattering is stronger in albite, since the scattering strength in stochastic region is proportional to scattering factor $Q''_{LL}$[16] (For albite $Q''_{LL} = 1.67 \times 10^{-2}$ while in iron $Q''_{LL} = 1.88 \times 10^{-3}$).

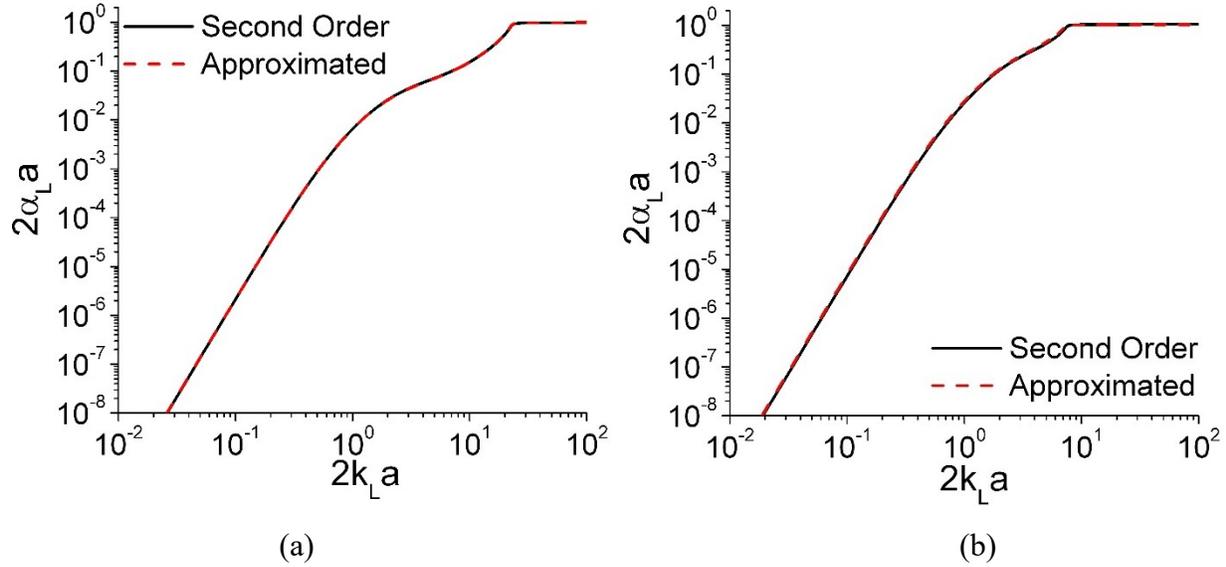

Figure 3 Longitudinal attenuation comparison between SOA (second order) model and ASOA (approximated) model for two polycrystalline materials (a) cubic iron (b) triclinic albite.

The discrepancy of longitudinal phase velocity between second order attenuation model and approximated model is evaluated in Figure 4, where the relative difference referred to Voigt velocity versus frequency constant is plotted. Again, two models are in good agreement. Further examination of Figure 4 shows that the maximum relative difference between two models is 0.04% on iron and 0.7% on albite, respectively. The discrepancy is attributed to the operation that inner product is factored out the from the double integration. Even large discrepancy happens on albite due to its stronger scattering strength.

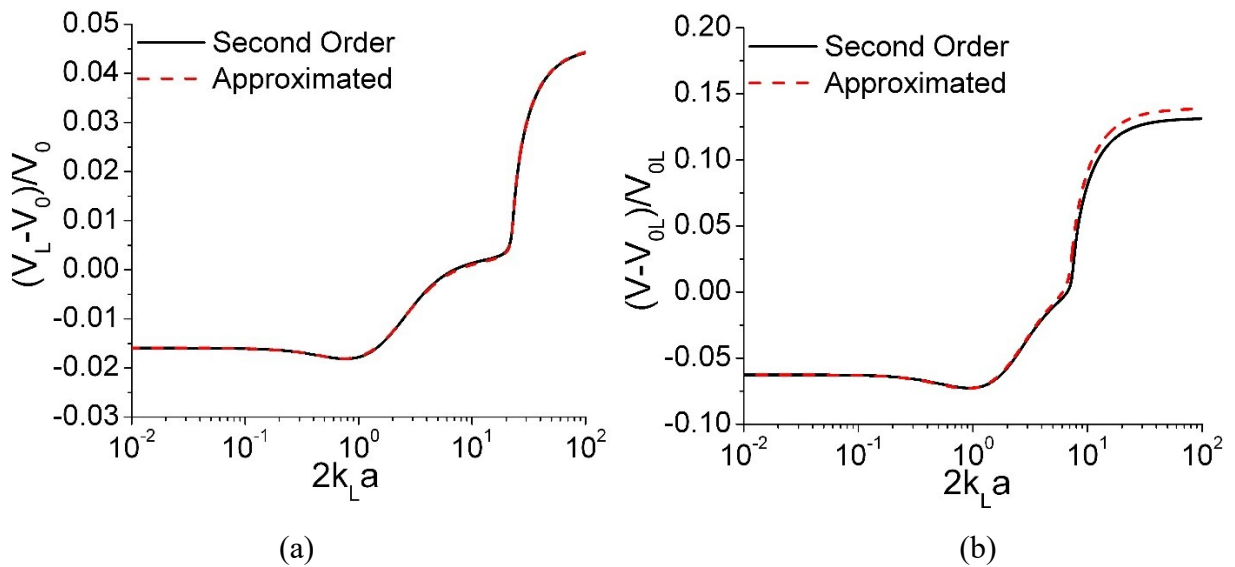

Figure 4 Longitudinal phase velocity comparison between SOA (second order) model and ASOA (approximated) model for two polycrystalline materials (a) cubic iron (b) triclinic albite.

Therefore, this section reports the results of attenuation and phase velocity obtained from SOA model for a triclinic polycrystal, which has never been achieved before. Also, the comparisons indicate the ASOA model is reasonable close to SOA model and has high efficiency and adequate accuracy, which is better than the far field scattering model in Ref.[16] where additional corrections of phase velocity are necessary.

**4.3 Comparison of ASOA with Karal&Keller Model on a hexagonal polycrystal**

Since the ASOA model has the form similar to Karal&Keller model [4], [5], it is also interesting to compare ASOA model with Karal&Keller model [4], [5]. However, for elastic waves, the inhomogeneity parameters in Karal&Keller model [4] have to be well defined. Ref.[5] explicitly defined the inhomogeneity parameters for polycrystals aggregates of hexagonal grains. Following the definitions for inhomogeneity parameters in Ref.[5], the Karal&Keller model for polycrystals aggregate consisted of equiaxed grains could be rewritten as:

$$k^2 = k_{0L}^2 + Q_{LL}^{''} k_{0L}^2 + \frac{4Q_{LL}^{''} k_{0L}^4}{k^2 - (k_{0L} + \frac{i}{a})^2} \quad \text{(longitudinal wave)} \quad (25a)$$

$$k^2 = k_{0T}^2 + Q_{TT}^{''} k_{0T}^2 + \frac{4Q_{TT}^{''} k_{0T}^4}{k^2 - (k_{0T} + \frac{i}{a})^2} \quad \text{(transverse wave)} \quad (25b)$$

where $Q_{LL}^{''}$ and $Q_{TT}^{''}$ have been given in section 3.1. Note dispersion equations are applicable to triclinic polycrystalline materials as well.

Titanium polycrystals aggregate [5] is used as an example to evaluate different models. Since second order attenuation model is available in this study, the attenuation and phase velocity from SOA are treated as reference. After calculations from different models are accomplished, the comparisons for longitudinal wave and transverse wave are presented in Figure 5 and Figure 6, respectively.

In Figure 5a the longitudinal attenuation curve from approximated model is in reasonable agreement with second order attenuation model while that from Karal&Keller model has a larger discrepancy (relative difference -76.7% in Table 1) at Rayleigh region ($2ka \ll 1$) and transition region ($2ka$ around 1). The large discrepancy in Karal&Keller model results from the missed L-T scattering mode, which is a dominant scattering portion at low frequency[28]. In Figure 5b, the phase velocity from Karal&Keller model is in poor agreement with second order attenuation model except stochastic region. Besides, the phase velocity in geometric region from Karal&Keller model goes to a different branch from second order attenuation model. Approximated model is much better than Karal&Keller model both on longitudinal attenuation and phase velocity.

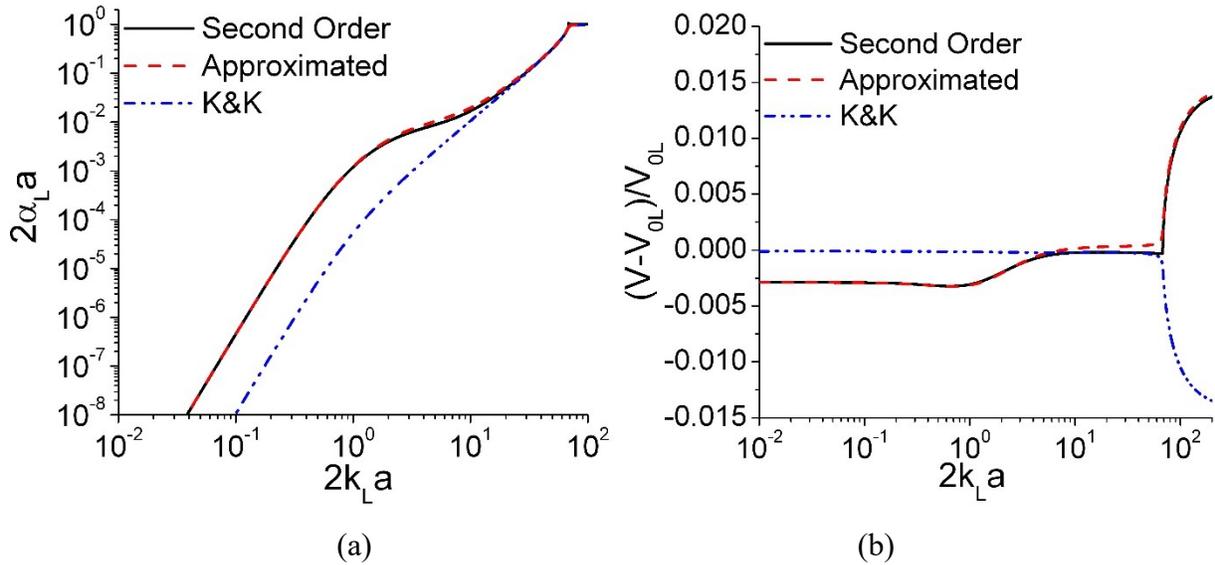

(a) (b)

Figure 5 Comparison of longitudinal attenuation and phase velocity between approximated attenuation model and Karal&Keller model [5] for titanium polycrystals (second order attenuation model is the reference): (a) longitudinal attenuation and (b) phase velocity. Only dominant root is shown for each model.

As for transverse attenuation, in Figure 6a the attenuation from ASOA model has good agreement with second order attenuation. Karal&Keller model also in good agreement with second order attenuation, and it is slightly better than ASOA model (see Table 1) since ASOA model somehow overestimates the attenuation coefficient in Rayleigh region and transition region. Unlike longitudinal wave case, the Karal&Keller model works well for transverse wave because T-T mode scattering is always dominant such that T-L mode scattering is negligible[28]. In Figure 6b, the phase velocity from ASOA model is reasonably close to the second order attenuation model while Karal&Keller model has a constant gap from second order attenuation model. Overall, ASOA model is much better than Karal&Keller model.

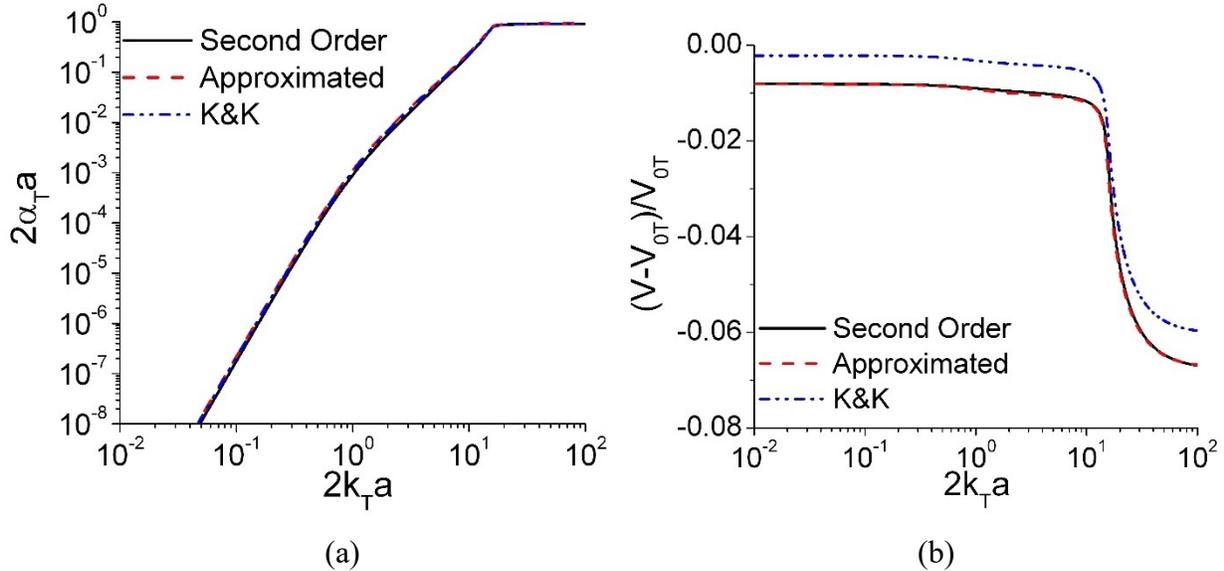

(a)            (b)

Figure 6 Comparison of transverse attenuation and phase velocity between approximated attenuation model (ASOA) and Karal&Keller model [5] for titanium polycrystals (second order attenuation model is treated as reference): (a) transverse attenuation and (b) phase velocity. Only dominant root is shown for each model.

Table 1 Mean relative difference of different models when compared with SOA model

|  | Longitudinal attenuation | Transverse attenuation |
| --- | --- | --- |
| ASOA | 18% | 23% |
| Karal&Keller model | 76.7% | 18% |

## 5. Conclusion

This paper developed a general second order attenuation (SOA) model for macroscopic isotropic polycrystals aggregates of triclinic equiaxed grains. Another approximated model, ASOA, is developed to simplify the second order attenuation model. Furthermore, comparisons of different models are addressed in this work. Several conclusions can be drawn:

1) The SOA model is applicable to triclinic polycrystalline materials without preferred crystallographic orientations and it is valid for whole frequency range. The development will allow the evaluation of the limitation of the Born approximation. The second order attenuation model is equivalent to well-known Stanke&Kino model[6] when applied to cubic polycrystals with equiaxed grains.

2) The approximated model, ASOAM, has high computation efficiency and adequate accuracy. It provides an efficient tool to predict elastic wave attenuation and velocity dispersion in polycrystalline materials

3) The ASOA model has better performance than Karal&Keller model[5] and can help us better understand the wave phenomena in polycrystalline media.

The models in this work can be potentially applied to the ultrasonic characterization of low symmetry grains, although further technical development is needed. This work also can be extended to texture-free polycrystals with ellipsoidal grains and textured polycrystalline materials in the future.